\documentclass[%
 aip,
 amsmath,amssymb,
 reprint,%
]{revtex4-1}

\usepackage{graphicx}
\usepackage{dcolumn}
\usepackage{bm}

\usepackage[utf8]{inputenc}
\usepackage[T1]{fontenc}
\usepackage{mathptmx}
\usepackage{etoolbox}
\usepackage{siunitx}
\usepackage{xr-hyper}
\usepackage[markup=underlined]{changes}

\newcommand \JK [1] {\bgroup\noindent[\textcolor{blue}{\textbf{JK}: #1}]\egroup\ignorespacesafterend}
\newcommand \SH [1] {\bgroup\noindent[\textcolor{red}{\textbf{SH}: #1}]\egroup\ignorespacesafterend}

\makeatletter
\def\@email#1#2{%
 \endgroup
 \patchcmd{\titleblock@produce}
  {\frontmatter@RRAPformat}
  {\frontmatter@RRAPformat{\produce@RRAP{*#1\href{mailto:#2}{#2}}}\frontmatter@RRAPformat}
  {}{}
}%
\makeatother
\begin{document}

\preprint{AIP/123-QED}

\title[X-ray particle tracking velocimetry for steady-state rheological characterization]{X-ray particle tracking velocimetry for steady-state rheological characterization: Case study of a complex polymer melt flow in material extrusion additive manufacturing}
\author{J. Kattinger}%
\affiliation{%
Institut für Kunststofftechnik, University of Stuttgart, Pfaffenwaldring 32, 70569 Stuttgart, Germany
}%

\author{S. Hiemer}%
\affiliation{Center for Complexity and Biosystems, Department of Physics "Aldo Pontremoli", University of Milan, Via Celoria 16, 20133 Milano, Italy
}%
\affiliation{%
CNR -- Consiglio Nazionale delle Ricerche, Istituto di Chimica della Materia Condensata e di Tecnologie per l'Energia, Via R. Cozzi 53, 20125 Milano, Italy
}%

\author{M. Kornely}%
\affiliation{%
Institut für Kunststofftechnik, University of Stuttgart, Pfaffenwaldring 32, 70569 Stuttgart, Germany
}%

\author{J. Ehrler}%
\author{P.-L. Chung}%
\author{C. Bonten}%
\author{M. Kreutzbruck}%
\affiliation{%
Institut für Kunststofftechnik, University of Stuttgart, Pfaffenwaldring 32, 70569 Stuttgart, Germany
}%

\begin{abstract}
We introduce X-ray Particle Tracking Velocimetry (XPTV) as a promising method to quantitatively resolve the velocity field and associated rheological information of polymer melt flow within the nozzle of a fused filament fabrication (FFF) printer. Employing tungsten powder as tracer particles embedded within a polymer filament, we investigate melt flow dynamics through an aluminum nozzle in a custom setup comparable to commercial printers. The velocity profiles obtained via XPTV reveal significant deviations from classical Newtonian flow, highlighting complex heterogeneous and non-isothermal behavior within the melt. From these measurements, we determine the local infinitesimal strain rate tensor and correlate flow-induced non-Newtonian effects to spatially varying temperature distributions, reflecting incomplete thermal homogenization within the nozzle. We complement the experiments with computational fluid dynamics simulations of the flow inside the printing nozzle, incorporating filament melting through an enthalpy–porosity formulation and treating the air–polymer melt interface using a two-phase approach. The simulated velocity profiles agree closely with the XPTV measurements across the investigated operating conditions, supporting the experimental interpretation. Our findings demonstrate the capability of XPTV to quantify both velocity fields and rheological properties, underscoring its potential as a tool for investigating opaque polymer melt flows in additive manufacturing, industrial processing, and rheology. To our knowledge, this is the first application of XPTV to polymer melt rheology. It enables measurements that are inaccessible to conventional optical methods.
\end{abstract}

\maketitle

\section{Introduction}
Standard rheological measurements such as in rotational or capillary rheometers rely on controlled flows with predictable kinematics to determine material properties. By calculating the stress and shear rate at specific points, for instance at the wall in a capillary rheometer, these methods infer rheological behavior without capturing the full flow profile \cite{macosko1994rheology}. In contrast, flow visualization techniques offer spatially resolved data, revealing detailed velocity profiles or stress distributions. This comprehensive approach has proven especially valuable in methods like low viscosity extensional rheometry, where direct imaging reveals the time-evolution of the diameter of a capillary \cite{bazilevsky1990liquid} or liquid jet \cite{schummer1983new,keshavarz2015studying} from which material properties are inferred \cite{mckinley2000extract}. Moreover, these techniques facilitate the analysis of complex flows, including heterogeneous, transient, or non-ideal conditions, commonly encountered in industrial manufacturing processes \cite{heindel2011review,poelma2020measurement,aliseda2021x}. 

A flow visualization technique with a long history is birefringent optics. It has been widely applied to non-Newtonian fluids, including Couette flow \cite{thurston_relaxation_1968, janeschitz-kriegl_polymer_1983, osaki_flow_1979}, parallel-plate and cone-plate flows \cite{10.1007/BFb0051073, janeschitz-kriegl_polymer_1983}, as well as capillary slit flows \cite{janeschitz-kriegl_polymer_1983, white_flow_1988}. By using the alignment between the stress tensor and the refractive index tensor, birefringence reveals stress contour lines and contributes to a deeper understanding of complex flow behaviors. Another established method, streak photography, uses tracer particles to visualize streamline patterns and is particularly valuable for studying entry flows \cite{binding_interfacial_1987, white_flow_1988}. Over the past decades, several optical techniques have been developed to study complex flow fields. Among them, laser Doppler velocimetry has become a method for pointwise velocity measurements by analyzing frequency shifts in scattered laser light caused by particles crossing the interference fringes of intersecting laser beams\cite{yeh_localized_1964, durst_principles_1976}. More comprehensive techniques such as particle image velocimetry (PIV) \cite{willert_digital_1991, pakdel_digital_1997, lee_x-ray_2003} and particle tracking velocimetry (PTV) \cite{adamczyk19882} have emerged, enabling full-field velocity mapping in two and three dimensions. PIV maps velocity fields by cross-correlating sequential particle image pairs, while PTV tracks individual particles, offering the added advantage of capturing the motion of individual particles over time, which reveals the time evolution of individual fluid elements within the flow field \cite{im_particle_2007}.

All of these techniques require optical access to the fluid, limiting their applicability to systems transparent for visible light. Alternative approaches such as radioactive particle tracking, nuclear magnetic resonance (NMR) \cite{xia1991study,callaghan1999rheo,xiong2024rheo} and adaptations of PIV and PTV using X-rays (referred to as XPIV and XPTV, respectively) have been used for studying optically inaccessible flows. Applications of XPIV and XPTV include investigations of flows in opaque circular pipes \cite{lee_x-ray_2003, fouras_three-dimensional_2007, im_particle_2007}, as well as more complex geometries such as bubble columns \cite{parker2024lab} and blood flow models \cite{kertzscher_x-ray_2004}. For radially symmetric flows, XPTV has been demonstrated using laboratory-scale X-ray sources at frame rates up to \SI{1}{\kHz} \cite{parker2024lab}.  Additionally, proof-of-concept studies highlight the feasibility of capturing full 3D flow fields using multiple source–detector pairs \cite{heindel_x-ray_2008}, tomographic techniques \cite{dubsky_computed_2012, makiharju2022tomographic}, or X-ray multi-projection imaging, which splits a single X-ray pulse into angularly separated beams \cite{rosen_synchrotron_2024}.  

\begin{figure}
\includegraphics[scale=0.4]{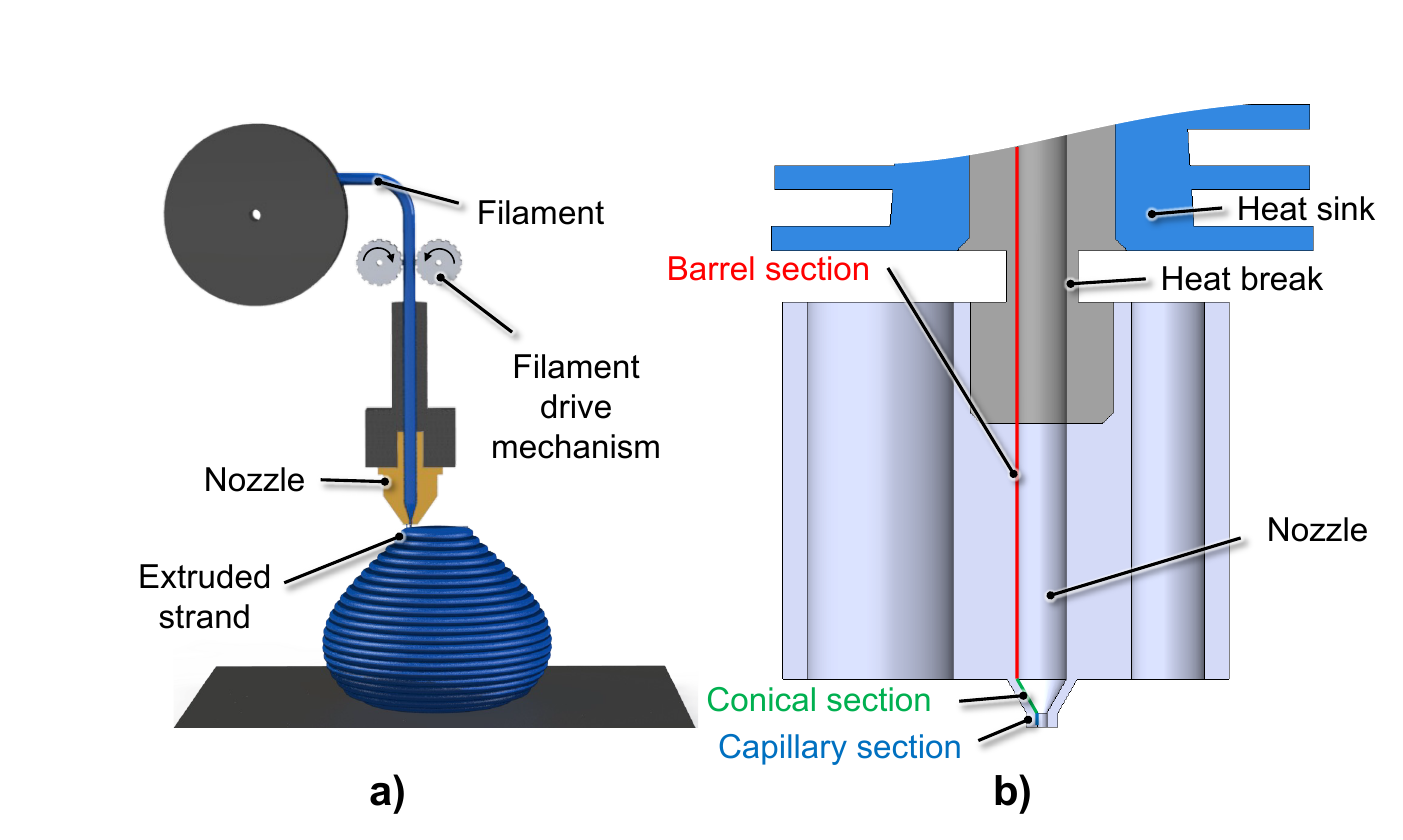}
\caption{\label{schematic_melting}Schematic illustrating (a) the FFF process, (b) the design of a typical nozzle and its connection to components such as the heater block, heat break, and heat sink.}
\end{figure}

Surprisingly, polymer melt flow—governed by complex rheological phenomena such as shear thinning as well as viscoelasticity and their complex property-structure relationship—has never been investigated using XPTV, despite its clear potential for studying opaque materials and manufacturing processes. One promising proof-of-concept application is fused filament fabrication (FFF), an extrusion-based additive manufacturing technique widely used to produce complex and highly functional polymer parts. As illustrated in Fig.~\ref{schematic_melting}(a), the process involves feeding a solid thermoplastic filament into a heated nozzle, where it melts and is extruded as a thin strand, layer by layer, onto a build platform to form a 3D object. Typically, a printing nozzle is composed of three distinct sections: the barrel, a conical transition zone, and the capillary (depicted in Fig.~\ref{schematic_melting}(b)). The filament, slightly smaller in diameter than the barrel, enters in its solid state and gradually melts through these sections. While this process may seem straightforward, the interplay of shear flows, thermal gradients, and material rheology creates significant challenges in controlling melting and flow dynamics within the nozzle \cite{mackay_importance_2018}. 

To gain a deeper understanding of the process, numerous modeling and experimental studies have been conducted. Analytical models vary widely in their assumptions, from homogeneous thermal conditions \cite{bellini_liquefier_2004, n_turner_review_2014} to those incorporating heat transfer to describe extremal cases of high print speeds \cite{osswald_fused_2018, colon_quintana_implementation_2020, luo_upper_2020, phan_rheological_2018} or minimal extrusion temperatures \cite{mackay_performance_2017, mackay_importance_2018}. With the exception of isothermal models, all analytical approaches emphasize the critical role of heat transfer in limiting print speed.

Numerical simulations have likewise been employed to address a broad range of questions. A central topic is die swell, which has been extensively studied using viscoelastic models \cite{de_rosa_experimental_2023,van_waeleghem_melt_2022, comminal_numerical_2018}. These typically focus on the flow immediately downstream of the nozzle exit under isothermal conditions. By contrast, simulations that describe filament melting within the nozzle solve an energy transport problem and generally treat the material as a generalized Newtonian fluid \cite{phan_computational_2020,pigeonneau_heating_2020, nzebuka_numerical_2022, marion_first_2023}. There is, however, one exception that solves the temperature equation while modeling the flow as viscoelastic \cite{xu_numerical_2024}. Another important distinction lies in the domain representation: either as a single phase \cite{pigeonneau_heating_2020, nzebuka_numerical_2022, ufodike_investigation_2021, kattinger_numerical_2022} or with a free-surface formulation to account for the air gap arising from the diameter mismatch between the filament and the barrel section \cite{serdeczny_numerical_2020, marion_first_2023}. Non-isothermal models consistently predict that incomplete melting can be significant and is strongly influenced by printing speed \cite{phan_computational_2020, serdeczny_numerical_2020, kattinger_numerical_2022}. Experimental measurements of the extrudate temperature at the nozzle exit further demonstrate that temperature gradients persist even at this stage \cite{zhang_temperature_2023}. This underscores the limited ability of the hot-end to fully heat the filament and raises fundamental questions regarding the actual flow and thermal states inside the nozzle.

Experimental studies can be broadly grouped into investigations of nozzle-exit phenomena and those probing the in-nozzle flow properties. Examples of the former include die swell and the temperature distribution at the nozzle exit \cite{colon_characterization_2023}. While such measurements are highly relevant for understanding print quality and process–property relationships, a complete process picture requires characterizing the flow inside the nozzle.

Pressure or temperature measurements within the nozzle \cite{anderegg_-situ_2019, coogan_-line_2019}, as well as feeding force measurements, are common approaches to infer in-nozzle behavior and assess model predictions \cite{serdeczny_experimental_2020, serdeczny_numerical_2020, kattinger_numerical_2022}. These setups offer crucial insights into the relationship between pressure loss and filament velocity and can even link the data to rheological material properties \cite{coogan_-line_2019}. However, like capillary rheometers, they infer flow properties solely from pressure measurements, without capturing the complete flow profile. Likewise, thermocouple measurements are intrinsically local and provide only an incomplete view of the temperature field, which is expected to vary with radius \cite{serdeczny_numerical_2020}.

To gain more detailed insights, one experimental study introduced dye markers into the filament to track their distribution after passage through the nozzle, revealing valuable information on the temperature history of the melt, namely that increasing filament velocity leads to more strongly non-isothermal conditions \cite{peng_complex_2018}. In another approach, a technique was developed that employs a glass nozzle in conjunction with a segmentally colored filament, enabling in-situ optical observation of the melt flow dynamics \cite{hong_-situ_2022}. Despite these advances, limitations remain—either due to the techniques not being in-situ or arising from thermal mismatches when using glass nozzles instead of metal ones. In our recent study, we employed in-situ X-ray computed tomography (CT) to investigate an aluminum printer nozzle operating under steady-state conditions \cite{kattinger2023analysis}. By resolving the polymer–air interface of the continuously fed filament, we revealed that increasing the filament velocity reduces the contact area between the melt and the nozzle wall. Furthermore, by introducing tungsten powder as a contrast agent into the filament and alternating its extrusion with contrast-free filament, we were able to track the interface between the two phases as temporal markers. While this experimental method provided valuable insights into the melting mechanism, it cannot quantify the underlying hydrodynamics, including velocity fields and shear distributions within the melt, which are essential for a comprehensive understanding of the flow behavior.

We present the application of X-ray Particle Tracking Velocimetry (XPTV) as a proof-of-concept technique to resolve the flow dynamics of a polymer melt flowing within the nozzle of a fused filament fabrication (FFF) printer. Building on Ref.~\onlinecite{kattinger2023analysis}, we now obtain spatially resolved velocity fields under the same process conditions by localizing individual tungsten tracer particles directly in the radiographs, rather than relying on contrast between particle-free and particle-laden segments. Our approach involves incorporating tungsten powder as tracer particles into a polymer filament and analyzing its flow through an aluminum nozzle integrated into a custom experimental setup. XPTV measurements reveal velocity profiles that clearly deviate from Newtonian flow, highlighting heterogeneous, non-isothermal conditions within the nozzle. By evaluating these velocity profiles, we estimate local strain rates and infer incomplete heating effects contributing to observed non-Newtonian behaviors. Additionally, we discuss the potential for reconstructing temperature distributions from flow data. To interpret these measurements, we complement XPTV with computational fluid dynamics simulations as a plausibility check for the observed trends. To our knowledge, this work represents the first use of XPTV to study polymer melt flow dynamics, demonstrating its promise as both a visualization and quantitative analytical method for opaque polymer systems commonly encountered in industrial processes.

\section{\label{chapter2}Materials and methods}
\subsection{\label{Materials}Materials}
This study examines the flow behavior of impact-modified polystyrene (PS486N, INEOS Styrolution, Germany) as it transitions from a solid filament to a melt within a heated aluminum nozzle. We used tungsten powder (W GG, H.C. Starck Tungsten GmbH, Goslar, Germany) as tracer particles due to its high X-ray attenuation. The tungsten tracers had an average particle size of \SI{32}{\um} (measured by laser diffraction), which allowed individual detection in X-ray radiography. We select the particle size distribution and concentration based on the specifications of our X-ray imaging setup, as detailed in section~\ref{Exp.setup}. The particles were incorporated into the polymer matrix via twin-screw extrusion at a concentration of 0.1 vol\%, optimizing particle tracking. The final compound was then extruded into a \SI{1.75}{\mm} diameter filament using a single-screw extruder. Further details on both compounding and extrusion are provided in Supplementary Information~S1.

\subsubsection{\label{Rheo}Rheological characterization}
\begin{figure}[h]
	\centering
		\includegraphics[scale=1]{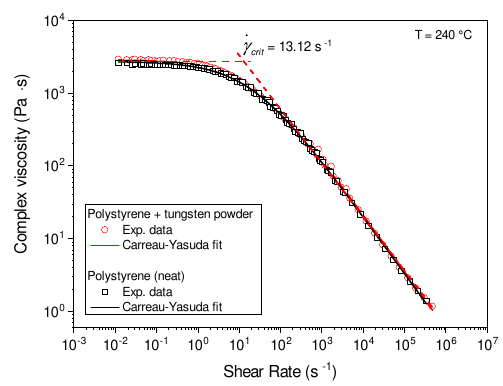}
	\caption{Master curves comparing complex viscosity as a function of shear rate for particle-laden and particle-free filaments.}\protect
	\label{fig:master}
\end{figure}

We characterized the rheological properties of the produced filament using a Discovery HR-3 rheometer (TA Instruments, New Castle, USA). To evaluate how tracer particles influence rheological behavior, we compared particle-laden filaments directly with particle-free filaments. We performed measurements using a \SI{25}{\mm} parallel-plate geometry and a \SI{1}{\mm} gap. Frequency sweep tests covered the range from \SIrange{0.01}{628}{\radian\per\second} at 2\% strain, ensuring conditions within the linear viscoelastic region typical for polymers \cite{osswald_polymer_2015}. Tests occurred at \SI{150}{\celsius}, \SI{205}{\celsius}, \SI{220}{\celsius}, and \SI{235}{\celsius} to explore temperature-dependent rheological behavior.
The magnitude of the complex viscosity \(|\eta^{\ast}|\) was converted to an apparent steady-shear viscosity using the Cox--Merz rule, \(\eta(\dot{\gamma}=\omega)=|\eta^{\ast}(\omega)|\) \cite{cox_correlation_1958}. For the tracer-laden material we verified the Cox–Merz correspondence by comparing frequency sweeps with steady-shear data in the range \SIrange{0.001}{1.0}{\per\second}. Finally, we applied Time-Temperature Superposition (TTS) using the Williams-Landel-Ferry (WLF) equation
\begin{equation}
\log a_T = -\frac{C_1 \left( T - T_0 \right)}{C_2 + \left( T - T_0 \right)}, \label{eq:at}
\end{equation}
and constructed master curves at \SI{240}{\celsius}. \(a_T\) is the shift factor, \( C_{1} \) and \( C_{2} \) are the parameters of the WLF equation, \(T\) the temperature and \(T_0\) the reference temperature. Fig.~\ref{fig:master} shows the resulting master curves. We fitted these curves using the Carreau-Yasuda model given by
\begin{equation} \label{eq: Carreau}
\frac{\eta}{\eta_0}=\left(1+\left(\lambda\dot{\gamma}\right)^{n_{1}}\right)^\frac{n_{2}-1}{n_{1}},
\end{equation}
where \( \eta_{0} \) is the zero-shear viscosity, \( \lambda \) is a characteristic time constant, and \( n_{1} \), \( n_{2} \) are model parameters. The critical shear rate \( \dot{\gamma}_{c} \) marking the transition from Newtonian to non-Newtonian behavior was estimated by determining the intersection of the power-law model, \( \eta = K \dot{\gamma}^{n_{3}} \) with power-law exponent \(n_3\) and flow consistency \(K\), and a zero-order model representing the zero-shear viscosity range. All corresponding model parameters for particle-laden and particle-free samples are detailed in Tab.~\ref{Rheo_parameters}. Additional details are provided in Supplementary Information~S2. 

\begin{table}[h]
\caption{Rheological Model Parameters}
    \label{Rheo_parameters}
\centering
\begin{ruledtabular}
\begin{tabular}{lcc}
Parameter & Particle-Free & Particle-Filled \\
\hline
\( C_1 \) (-) & 4.6  & 4.4  \\
\( C_2 \) (K)\ & 200.2  & 196.8  \\
\( T_0 \) (°C) & 220  & 220 \\
\( \eta_0 \) (Pa·s)  & 2751  & 2872 \\
\( \lambda \) (s) & 0.052 & 0.055 \\
\( n_1 \) (-) & 0.62 & 0.59 \\
\( n_2 \) (-) & 0.22 & 0.22 \\
\( n_3 \) (-) & 0.26 & 0.26 \\
\( K \) (-) & 15983  & 16068  \\
\( \dot{\gamma}_{c} \) (1/s) & 12.4  & 13.1  \\
\end{tabular}
\end{ruledtabular}
\end{table}

To quantify the viscosity increase due to tracer particles, we applied the Einstein viscosity relation
\begin{equation}
\eta_{0} = \eta_{0,\text{neat}} \left(1 + 2.5 \phi\right),
\end{equation}
which predicts a 0.25\% increase in zero-shear viscosity. However, the Carreau–Yasuda fits indicate an increase of approximately 4\% due to the addition of the tracer particles. The higher viscosity likely results from the additional twin-screw extrusion of the tracer-filled compound, which may have changed its microstructure. As the flow examined here involves only the tracer-containing melt, the particle-free rheology does not enter the subsequent interpretation and is reported for completeness.

\subsubsection{\label{Thermal}Thermal characterization}
\begin{figure}[h]
	\centering
		\includegraphics[scale=1]{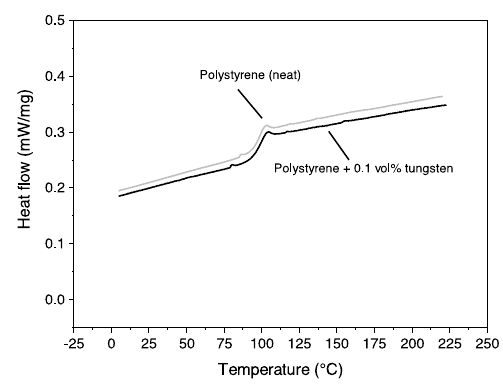}
    \caption{Differential Scanning Calorimetry (DSC) results for particle-laden and particle-free filaments, demonstrating that the addition of tungsten particles has no measurable effect on the glass transition temperature.}\protect
	\label{DSC} 
\end{figure}

We analyzed the thermal properties of the produced filament, both with and without tracer particles, using differential scanning calorimetry (DSC) on a Mettler Toledo DSC 2 instrument (Mettler-Toledo, LLC, Columbus, USA). A small sample, sealed in an aluminum pan, was heated at a rate of \SI[per-mode=symbol]{10}{\K\per\minute} from \SI{20}{\celsius} to \SI{250}{\celsius}. Following a 5 min isothermal phase at \SI{250}{\celsius}, the sample was cooled back to \SI{20}{\celsius} at the same rate. As shown in Fig.~\ref{DSC}, the addition of tungsten powder caused a vertical shift to lower values in the DSC curve, attributed to a reduction in the heat capacity of the material. To avoid any potential influence of thermal history, the second heating curve is typically considered because it provides more consistent and reliable data regarding intrinsic thermal transitions of the material. Consequently, the second heating curve revealed a glass transition temperature, $T_g$, of \SI{101}{\celsius}, determined by the midpoint method, which remained unaffected by the tungsten powder addition. These results confirm that adding tungsten powder predictably modifies the rheological and thermal properties while maintaining the fundamental characteristics of the neat polymer, thus effectively functioning as a contrast agent.

\subsection{\label{Exp.setup}Experimental setup and procedure}
\subsubsection{\label{Polymer_Extr}Polymer extrusion setup}
\begin{figure*}
\includegraphics[scale=0.8]{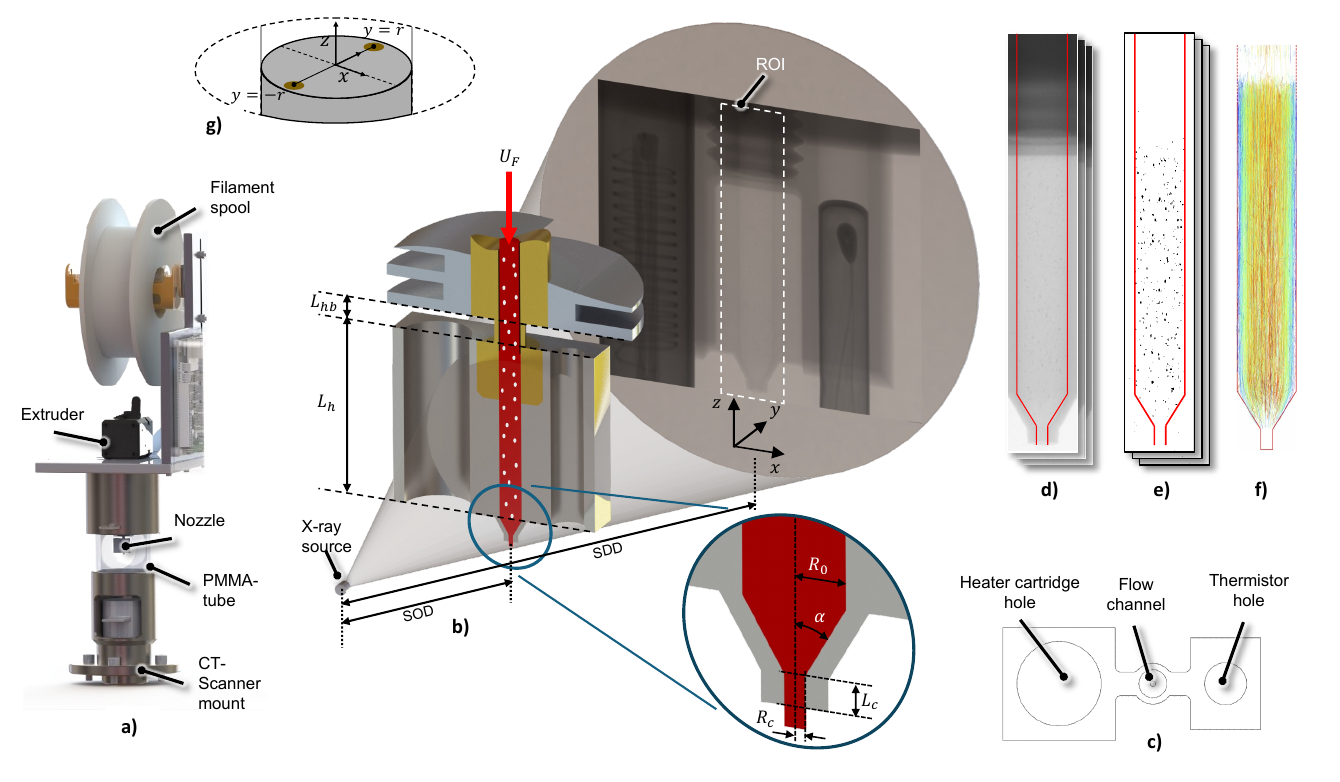}
\caption{\label{fig:schematic-hotend}Schematic of the polymer extrusion setup and the X-ray imaging principle for circular flow with tracer particles. (a) Overview of the extrusion setup, based on a previous version described in Ref.~\onlinecite{kattinger2023analysis}, adapted for integration with the X-ray system. (b) Sectional view of the manufactured nozzles, showing dimensions and orientation relative to the cone beam and detector. (c) Top view of the nozzle. (d) Cropped raw image stack of the region of interest (ROI), showing low particle contrast. (e) Background-subtracted image stack with enhanced contrast. (f) Particle tracking results with velocity-coded tracks. (g) Schematic representation of particles positioned at \(y = r\) and \(y = -r\).}
\end{figure*}

The experimental setup shown in Fig.~\ref{fig:schematic-hotend}(a), originally introduced in Ref.~\onlinecite{kattinger2023analysis}, was employed to study the flow behavior within a heated nozzle using 360°-CT measurements and 2D projectional radiography during continuous material extrusion, as found in FFF printers. The setup was designed for operation inside a CT scanner. A PMMA tube connects the upper and lower sections and ensures unobstructed radiography of the nozzle.

In this study, we utilized X-ray particle tracking velocimetry to analyze 2D radiography data captured through video recording. To enhance X-ray transparency and optimize the setup, we designed and fabricated a custom aluminum nozzle. Its flow channel dimensions (given in Fig.~\ref{fig:schematic-hotend}(b) and Tab.~\ref{Nozzle_geometry_parameters}) match those of commercially available printing nozzles, ensuring practical relevance. A key design criterion was maximizing contrast for the flowing tracer particles, achieved by minimizing wall thickness along the radiation beam direction, as seen in the top view of the nozzle in Fig.~\ref{fig:schematic-hotend}(c).

We used a commercial extruder (Bondtech LGX model, Värnamo, Sweden) to continuously feed the solid filament into the nozzle, which includes a heater cartridge and a thermistor. An E3D-V6 type hot-end connected the feeding mechanism to the custom nozzle. The experimental setup was controlled by a PC linked to a Duet 2 Ethernet board (Duet3D, Peterborough, United Kingdom) running RepRap firmware. Instead of depositing the strand on a build platform, we directed the extruded strand into a pan positioned below the nozzle orifice. 

\begin{table}
\caption{\label{Nozzle_geometry_parameters}Nozzle geometry parameters}
\begin{ruledtabular}
\begin{tabular}{ccddd}
Parameter & Value \\ \hline
$R_c$ (mm)  & 0.2   \\
$R_0$ (mm)  & 1.0   \\
$L_c$ (mm)  & 0.6   \\
$L_h$ (mm)  & 15.5   \\
$L_{hb}$ (mm) & 2.1 \\
$\alpha$ (deg) & 30   \\ 
$d$ (mm) & 0.2   \\ \hline
\end{tabular}
\end{ruledtabular}
\end{table}

\subsubsection{\label{X-ray}X-ray imaging setup}
The polymer extrusion setup was mounted in a YXLON FF20 ${\mu}$-CT Scanner, equipped with a Comet FXE190.61 transmission X-ray tube (Comet AG, Wünnewil-Flamatt, Switzerland) and a Varex 2530HE detector (Varex Imaging, Salt Lake City, Utah, USA), providing 16-bit image depth and a pixel pitch of \SI[per-mode=symbol]{139}{\micro\meter}. In this study, we utilized the ${\mu}$-CT Scanner to produce 2D data through projectional radiography. 

\subsubsection{Polymer extrusion parameters}
Our experiments focus on observing the flow inside a heated printer nozzle as the filament is fed at four different velocities: \SI[per-mode=symbol]{0.5}{\mm\per\s}, \SI[per-mode=symbol]{1.0}{\mm\per\s}, \SI[per-mode=symbol]{1.5}{\mm\per\s}, and \SI[per-mode=symbol]{2.0}{\mm\per\s}, while maintaining a constant heater temperature of \SI{240}{\celsius}. In addition, for the case of \SI[per-mode=symbol]{1.5}{\mm\per\s}, the extrusion process was also studied at heater temperatures of \SI{220}{\celsius} and \SI{260}{\celsius}. Prior to testing the different filament velocities, the feeding mechanism was calibrated to ensure the actual filament velocity matched the commanded velocity. After calibration, polymer extrusion was run for five minutes, with the first \SI{60}{\s} excluded from recording to capture only steady-state conditions.  
\subsubsection{X-ray imaging parameters}
The recording was conducted at a tube voltage of \SI{190}{\kV} and a current of \SI{60}{\uA}, delivering an X-ray system power output of \SI{11.4}{\W}. A 2×2 pixel binning was applied to enhance the signal-to-noise ratio and achieve a frame rate of \SI{15}{\Hz}. The corresponding exposure time was \SI{67}{\ms}, determined by the frame rate and set to its maximum feasible value to ensure sufficient contrast-to-noise ratio given the X-ray source and detector. This frame rate restricts analysis to the barrel section. In the conical region the flow accelerates so rapidly that some particles traverse the cone and capillary within a single frame.

The region of interest (ROI) was defined as the section of the nozzle marked by a white dashed line in Fig.~\ref{fig:schematic-hotend}(b). To optimally capture this ROI, the source-to-detector distance (SDD) and source-to-object distance (SOD) were adjusted, resulting in values of \SI{805}{\mm} and \SI{54}{\mm}, respectively. Due to the specific nature of conical X-ray sources, a geometric magnification results that is directly related to the distance from the X-ray source. Consequently, the magnification varies with depth throughout the experimental geometry. Similarly, the detected particle image movement is affected by magnification \(M\). Particles closer to the source appear larger and move more rapidly than those farther away, even if their size and velocity are identical. Calibration is conducted at the nozzle center, positioned at a distance SOD from the X-ray source. For particles located at \(y = -r\) and \(y = r\) on the pipe wall (as shown in Fig.~\ref{fig:schematic-hotend}(g)), the resulting magnification error due to their distance from the center is calculated as
\begin{equation}
\frac{\text{M}_{\text{(y)}}}{\text{M}_{\text{(y=0)}}} = \frac{\text{SOD}}{\text{SOD+y}}.
\end{equation}
In the experiments, we observe a magnification error of 1.3\,\%, meaning that the maximum deviation between the observed particle velocity and its actual movement never exceeds this value.

\subsubsection{\label{Particle}Particle size distribution and concentration}
\begin{figure}
	\centering
		\includegraphics[scale=0.9]{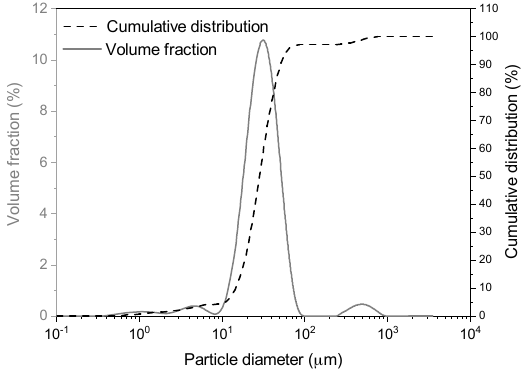}
	\caption{Particle size distribution obtained by laser diffraction, showing particle diameter on the x-axis and corresponding volume fraction and cumulative distribution on the y-axis.}\protect
	\label{Laser_difration}
\end{figure}

Accurate particle tracking depends on tracer particles that are sufficiently large for reliable detection. With an active pixel area of $\SI{278}{\micro\meter} \times \SI{278}{\micro\meter}$ resulting from binning and the selected SDD and SOD, the resolution achieved is \SI[per-mode=symbol]{18.5}{\um}, which defines the minimum detectable particle size. Consequently, a powder was chosen that, according to laser scattering analysis (Fig.~\ref{Laser_difration}), contains a sufficient number of particles within the desired size range ($d_{10}=\SI{15}{\um}$, $d_{50}=\SI{29}{\um}$, and $d_{90}=\SI{51}{\um}$).

Optimizing particle concentration was essential to balance ease of detection and recording efficiency. While lower concentrations simplify particle identification, they require longer recording durations for adequate flow data acquisition. Conversely, higher concentrations lead to increased particle overlap at different depths, complicating detection. An optimal average inter-particle spacing was defined as eight times the mean particle diameter, resulting in a separation distance of \SI{258}{\um}.  This factor ensures a sufficient number of detectable tracer particles while minimizing overlap in depth, which can hinder accurate detection in volumetric tracking methods. Assuming spherical particles, the particle volume \( V_{\text{Sphere}} = \frac{\pi}{6} d^3 \) relative to the unit cell volume \( V_{\text{UC}} = L^3 \), and an average particle diameter of \SI[per-mode=symbol]{32}{\um}, yields a tungsten particle volume fraction of 0.1\%.

\subsubsection{\label{Image_processing}Image processing}
Before computing tracer velocities, the raw images were pre-processed to enhance the detectability of each tracer particle. The processing steps leading to the calculation of tracer velocities and their trajectories were performed using the open-source software Fiji and are illustrated in Fig.~\ref{fig:schematic-hotend}(d-f). Initially, the raw images were cropped to isolate the ROI ($\SI{16.4}{\milli\meter} \times \SI{2.6}{\milli\meter}$). Given the low contrast in the original images, a time-averaged background image was generated and subtracted from each frame to improve particle visibility.

Next, particle tracking velocimetry was performed using the Fiji plugin TrackMate \cite{ershov_trackmate_2022}. This process involved two main steps: particle detection and particle tracking. For particle detection, the Laplacian of Gaussian (LoG) detector was employed, which detects particles by identifying local intensity maxima in the image after applying a Gaussian blur. On average, this resulted in approximately 250 detected spots per image, which were then tracked in the subsequent step using the Linear Assignment Problem (LAP) tracker. This method calculates particle trajectories by minimizing the overall cost of linking based on criteria such as proximity and predicted motion \cite{ershov_trackmate_2022}. Further details on the tracking parameters and associated artifacts are provided in Supplementary Information~S3.

To enable further analysis and visualization of particle paths, the data was processed in Python using the NumPy library \cite{harris2020array}. First, pixel-to-millimeter scaling was performed. Given that the pipe diameter is known to be \SI{2.0}{\mm}, the number of pixels spanning this diameter was counted to determine the scale.

The next step involved removing calculated particle paths that originate from faulty detections. Ideally, a calculated particle path results from tracking a particle through the entire flow channel. However, particle detection and tracking are imperfect: particles with low contrast-to-noise ratio may intermittently alternate between detection and non-detection as the LAP tracking algorithm advances through the frames. In some cases, rather than losing detection, a nearby particle may be mistakenly identified and combined into the same path. This results in a zig-zag pattern in the computed particle path. To address this issue, we deleted all tracks where the ratio of y-direction forward progression to the total path length was less than 80\%. This criterion primarily eliminates short paths with low-quality particle detections (as measured by quality metric of TrackMate), reducing the risk of introducing bias into the results. 

\subsection{\label{fidelity}Tracer particle fidelity via dimensional analysis}
The precision of our experiments depends significantly on the ability of tracer particles to accurately follow the flow dynamics. To ensure this, we evaluate the effects of gravity and inertia on particle movement. Gravity induces a settling velocity, \(u_{St}\), which quantifies the rate at which particles descend due to gravitational forces. The settling velocity is defined as 
\begin{equation}
u_{St} = \frac{d_{p}^2 (\rho_{p} - \rho_{f}) g}{18 \eta},
\end{equation}
where \(d_p\) is the particle diameter, \(\rho_p\) and \(\rho_f\) represent particle and fluid densities (density values of the materials used are summarized in Tab.~\ref{density_parameters}), respectively, \(g\) is gravitational acceleration, and \(\eta\) is the dynamic viscosity of the fluid. Given the non-Newtonian behavior of our fluid, we use the zero-shear viscosity \(\eta_0 = \SI{2872}{\pascal\second}\). Considering the known densities of both particles and fluid, the calculated settling velocity is \(u_{St} = \SI{4E-6}{\mm\per\second}\). Since this velocity is significantly smaller than the characteristic velocities of the fluid, which are of the order of the filament velocity, \(\sim \SI{1E0}{\mm\per\second}\), gravitational effects on particle trajectories are effectively negligible.

\begin{table}
\caption{\label{density_parameters}Density parameters}
\begin{ruledtabular}
\begin{tabular}{l c S[table-format=2.3] l}
Density property & Symbol & {Value} & Unit \\ \hline
Polymer fluid      & \(\rho_f\) & 0.92 & \si{\gram\per\cubic\centi\meter} \\
Polymer solid      & \(\rho_s\) & 1.02 & \si{\gram\per\cubic\centi\meter} \\
Tracer particle    & \(\rho_p\) & 19.3 & \si{\gram\per\cubic\centi\meter} \\
Density ratio (fluid to tracer) & \(\rho_p/\rho_f\) & 21 & dimensionless
\end{tabular}
\end{ruledtabular}
\end{table}

Another critical factor is the Stokes number (\(St\)), a dimensionless quantity that characterizes the response time of particles relative to a characteristic time of the flow. The Stokes number is given by
\begin{equation}
St = \frac{u_{c} t_{c}}{l_{c}},
\end{equation}
where \(u_{c}\), \(t_{c}\), and \(l_{c}\) denote the characteristic velocity, time, and length scales of the flow, respectively. A Stokes number much less than one (\(St \ll 1\)) suggests that particles closely follow the streamlines of the fluid. We define the characteristic length scale \(l_{c}\) as the diameter of the barrel section and set the characteristic velocity \( u_c \) to \SI{10.0}{\mm\per\s}, assuming the flow does not exceed this value anywhere in the nozzle, when the highest filament velocity of \SI{2.0}{\mm\per\s} is applied.

Under the assumption of a Reynolds number less than one, the characteristic time of a particle \(t_{c}\) can be approximated using Stokes flow assumptions
\begin{equation}
t_{c} = \frac {\rho _{p} d_{p}^{2}}{18 \eta _{0}}.
\end{equation}
We compute \( St = 1.6 \times 10^{-6} \), confirming that \( St \ll 1 \) and ensuring the tracer particles accurately track the fluid motion.  

\subsection{Numerical simulations}
As part of this work, we performed computational fluid dynamics (CFD) simulations using the commercial finite volume software ANSYS Fluent 2024 to resolve the velocity and temperature field inside the nozzle. We expect the temperature field to depend sensitively on the axial location at which the polymer first contacts the barrel wall and on the downstream extent of any residual air gap. To capture this behavior, the process is modeled as two-phase flow in a Volume of Fluid (VOF) framework, with polymer and air treated as immiscible phases that share momentum and energy equations.

\begin{figure}
	\centering
		\includegraphics[scale=0.45]{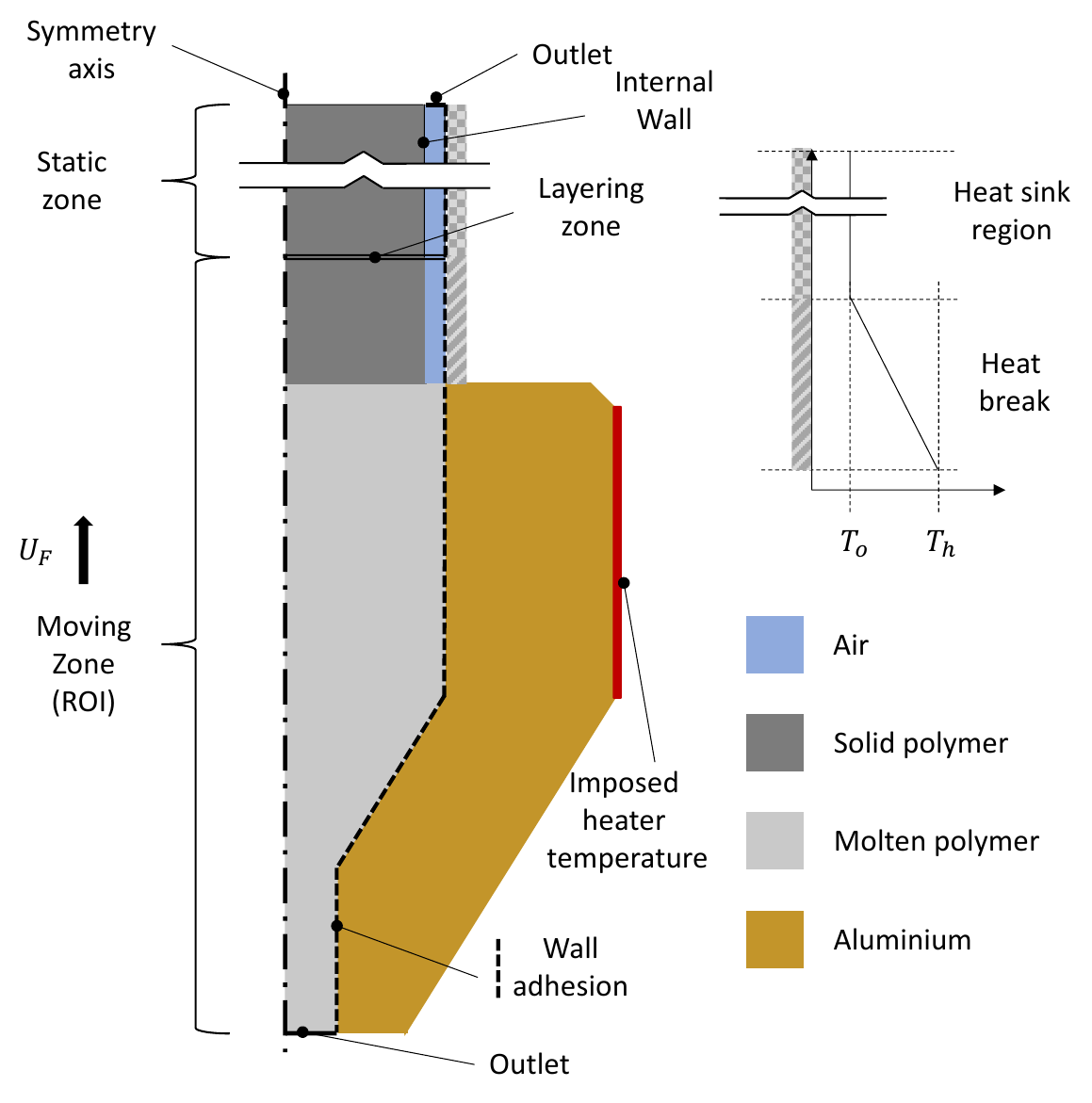}
	\caption{Numerical domain and boundary conditions of the axisymmetric nozzle model, including the applied initial condition. Polymer and air are resolved as immiscible phases using a VOF approach.}
\protect
	\label{Numerical_domain}
\end{figure}

The configuration (Fig.~\ref{Numerical_domain}) is two dimensional and axisymmetric about the nozzle axis. Conjugate heat transfer is solved in the metallic nozzle and in both fluids so that a non-uniform temperature distribution in the metal can feed back on the local melt temperature. Melting of the incoming filament is represented by the enthalpy–porosity technique, which introduces a Darcy-type momentum sink in partially molten cells and yields a continuous transition from solid-like to liquid-like behavior \cite{voller1987fixed}. This formulation, however, also damps the solid phase, which prevents prescribing a constant filament velocity. To overcome this, a dynamic mesh with axial layering is employed \cite{ANSYSFluent2024}. The domain is advanced relative to the filament at a prescribed speed, which is equivalent to feeding the filament through a fixed hot-end while preserving the solid-phase damping. The simulations are transient. At the initial time the nozzle fluid space is filled with polymer melt at the heater set temperature, after which cold filament at ambient temperature is pushed into the domain at the same four feed velocities as in the experiment.

Thermal and physical properties of the polymer are temperature dependent. The viscosity follows a Carreau–Yasuda law (Eq.~\ref{eq: Carreau}) with a WLF temperature shift (Eq.~\ref{eq:at}) implemented via a user-defined function calibrated from the rheology data in Section~\ref{Rheo}. The thermal conductivity \(k(T)\), the heat capacity \(c_p(T)\), and the density \(\rho_f\) of the polymer are modeled as functions of temperature (see Supplementary Information~S5 for details), while the metallic nozzle is assigned constant properties of aluminum, with \(k = 237 \,\text{W m}^{-1}\text{K}^{-1}\) and \(c_p = 900 \,\text{J kg}^{-1}\text{K}^{-1}\). Viscous heating was neglected, as the expected contribution of dissipative heating to the overall energy balance is small compared to conductive heat transfer from the nozzle walls. All internal solid walls satisfy the no-slip boundary condition. We represent the heat break by prescribing a linear decrease of the wall temperature from the heater set temperature to ambient. On the subsequent wall section inside the heat break we apply an ambient-temperature boundary condition. Upstream of the heat break, we model the heat-sink region as a reservoir of solid filament. In the moving-mesh formulation, the nozzle advances over this reservoir at the prescribed feed speed.

The domain employs two pressure outlets at ambient: i) the nozzle orifice, through which the extrudate and any entrained air leave the domain, and ii) an air-vent outlet at the upper end of the flow domain, which allows displaced air to escape and prevents artificial pressure build-up. On the outer wall of the nozzle adjacent to the heater block we prescribe the heater set temperature, while all other metal surfaces in contact with the polymer or air are thermally coupled. Details of solver formulation, discretization schemes, interface compression parameters, time-step control, and convergence thresholds are provided in the Supplementary Information~S5.

\section{\label{Results}Results and discussion}
\begin{figure*}
\includegraphics[scale=1.05]{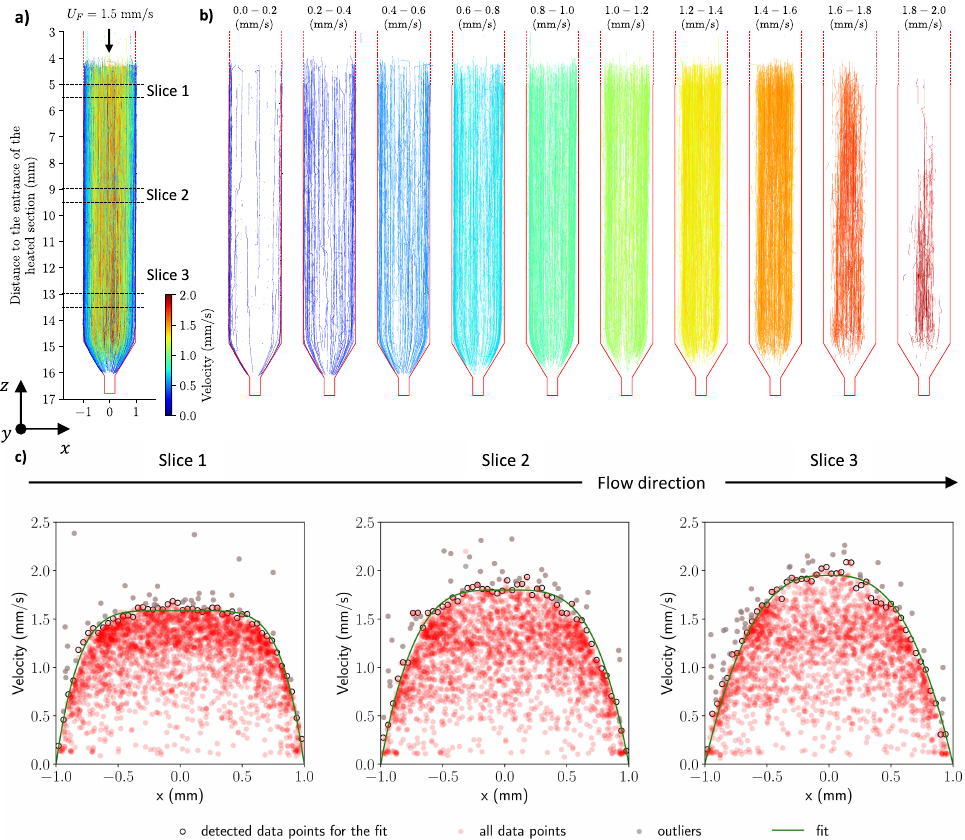}
\caption{\label{fig:particle_paths}Particle paths visualized using a velocity-based color code at a filament velocity of $U_F = \SI{1.5}{\mm\per\s}$: (a) all paths combined, with highlighted virtual slices; (b) paths grouped into ten ranges based on their mean velocity. Occasional lateral jumps are artifacts of projection and tracking (for example identity loss during gap closing) and should not be interpreted as physical cross-stream motion, which is expected to be small and below the spatial resolution of the present setup (see Supplementary Information~S3); (c) scatter plot showing the velocity and x-position of the particles as they pass the virtual slices.}
\end{figure*}

\subsection{Velocity and strain rate calculation}
The extracted signal of our measurements are projections of the particles onto the detector and along their respective paths. While the particle paths depicted in Fig.~\ref{fig:particle_paths}(a) and (b) offer qualitative insights, they must first be transformed into velocity profiles to enable the extraction of rheologically relevant information, such as strain rates.

To this end, we discretize the nozzle along the z-axis into axial slices of thickness \(\Delta z = \SI{0.5}{\milli\meter}\). For each slice we compute the mean velocity and mean radial position of every particle that passes during a fixed acquisition window of \(T = \SI{3}{\minute}\). The resulting ensemble is visualized as scatter plots (Fig.~\ref{fig:particle_paths}(c)) that show the distribution of particle velocities along the projection direction (y-axis). Assuming axisymmetry of the velocity field, the scatter plots provide the basis for recovering the centerline velocity distribution by extracting an envelope of the maximum radially observed particle velocities. In doing so, we map Lagrangian tracer trajectories to an Eulerian description by binning observations into axial slices of defined thickness and collecting them over the window \(T\), yielding an Eulerian field \(u_z(r,z)\) that is localized in \(z\) by \(\Delta z\) and constructed from a time-collected ensemble. Complete sets of scatter plots covering all axial positions along the flow, filament velocities, and temperatures are provided in Supplementary Information~S8.To extract these envelopes, we first applied the \mbox{DBSCAN} filter \cite{ester1996density} to remove isolated data points located in low density regions. A targeted inspection of the DBSCAN outliers showed that they arise when two particles are erroneously merged into a single tracking path, introducing a trajectory discontinuity (jump) that yields spuriously high velocities. Subsequently, we segment the data into 50 equally spaced radial sections and select the fastest velocity data point from each segment, excluding any points previously identified as outliers by the DBSCAN filter. Finally, the selected points are fitted to the function
\begin{equation}
    u_{z}(r) = a_{z} \left(1 - \left( \frac{r}{R} \right)^{b_{z}} \right), \label{eq:fitting}
\end{equation}
with fit parameters \(a_{z}\) and \(b_{z}\). To avoid confusion, we note that this fitting function is identical to the solution for an isothermal power-law fluid in Poiseuille flow. However, it was chosen solely for its ability to represent the data, its analytical convenience, and ease of interpretation, with no deeper rheological implications.

\begin{figure*}
	\centering
		\includegraphics[scale=0.795]{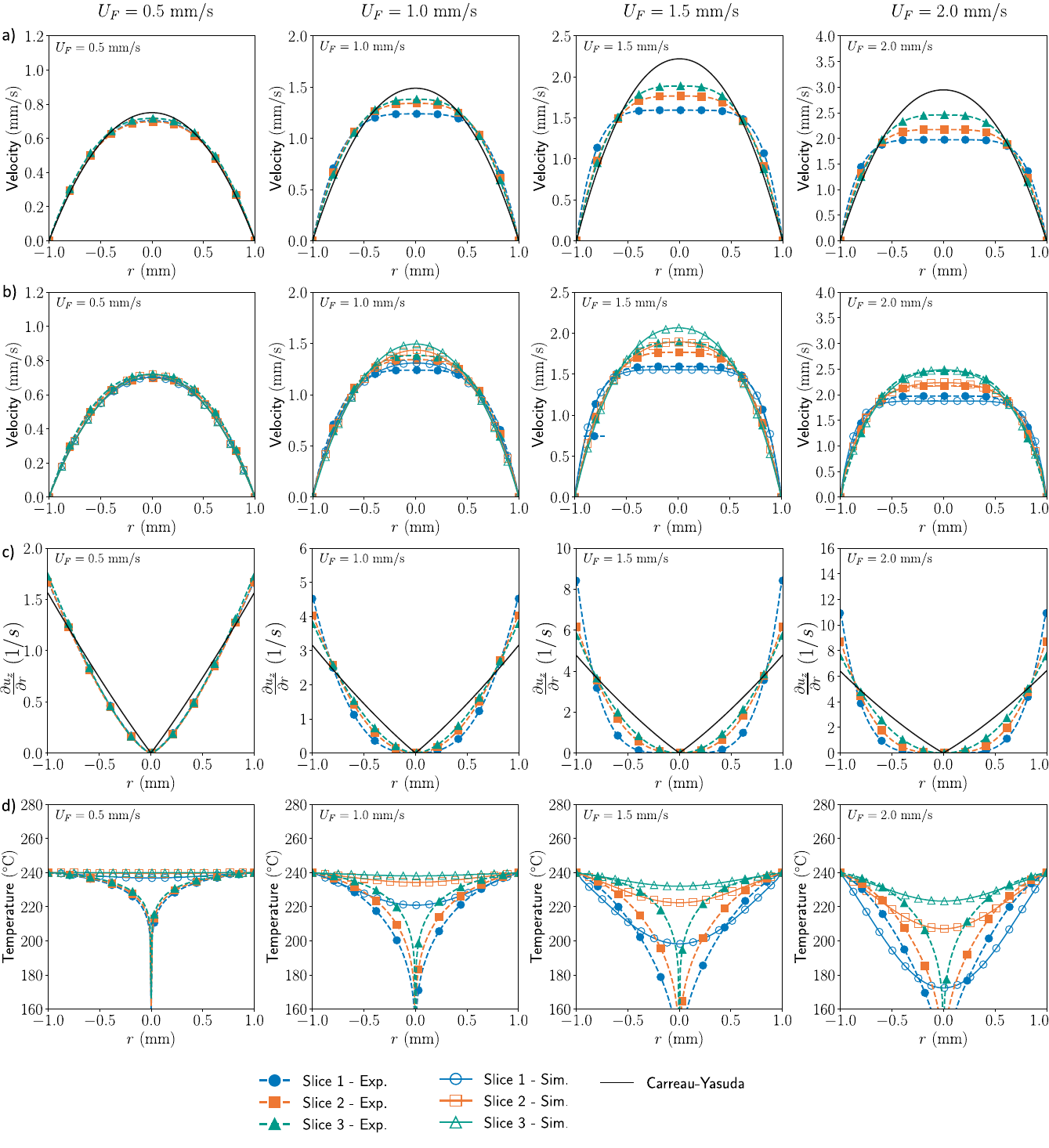}
	\caption{Matrix arranged by filament velocity along the column axis. (a) Comparison between experimentally obtained velocity profiles and the one-dimensional (1D) Carreau-Yasuda model representing isothermal conditions. (b) Comparison between experimentally obtained velocity profiles and CFD simulation results. (c) Shear-rate profiles derived from experimental velocity fields and compared with the 1D Carreau-Yasuda model. (d) Preliminary reconstruction of temperature distributions obtained from experimental data and compared with corresponding CFD predictions. Different colors indicate distinct axial positions along the nozzle.}\protect
	\label{fig:velocity_profile}
\end{figure*}

\begin{figure}
    \centering
    \includegraphics[scale=0.86]{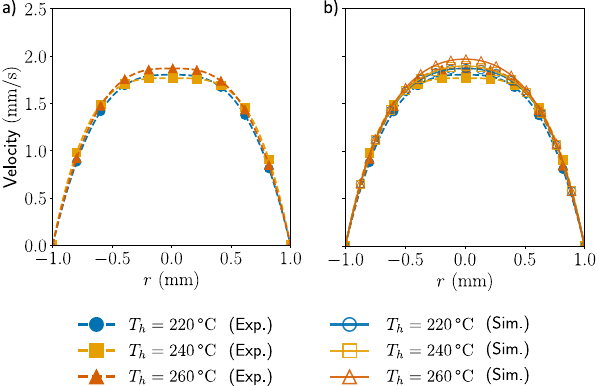}
    \caption{Velocity profiles at slice 2 for a filament velocity of 1.5 mm/s at different heater temperatures. (a) Experimental data. (b) Comparison between experimental and simulation results (different colors indicate different heater temperatures).}
    \label{fig:Influence_of_heater_temp}
\end{figure}

As demonstrated in Fig.~S5 both fit parameters vary approximately linearly along the flow direction, so \(u_{z}(r,z)\) is effectively known (see Appendix \ref{sec:vel-grad} for details). A closer examination of the velocity profile at three positions along the flow (Fig. \ref{fig:velocity_profile}(a) and (b)) or via the color-coded velocity plot (Fig.~S12) and the velocity gradients (Fig. \ref{fig:velocity_profile}(c)) reveals that the profile is nearly Newtonian at the lowest filament velocity. At higher filament velocities, however, the profile becomes increasingly plug-like and transitions  to a parabolic shape when moving closer to the conical section. Notably, varying the filament velocity alters the profile much more strongly than changing the heater temperature (Fig.~\ref{fig:Influence_of_heater_temp}). In fact, adjusting the heater setpoint between \SI{220}{\celsius} and \SI{260}{\celsius} yields velocity profiles that are barely distinguishable.

To further investigate the experimentally observed flow phenomena, we compared the measurements with two complementary numerical models. The first is a one-dimensional isothermal Poiseuille-flow solution for a Carreau-Yasuda fluid (see Supplementary Information~S4), which we used to test whether nonlinear effects in the transition from the Newtonian plateau to the power-law regime can explain the observations. The second consists of non-isothermal CFD simulations that solve the energy equation to model filament melting and serve as a plausibility check for the experimental results. While we observed close agreement between the isothermal numerical and experimental data at low filament velocities, significant deviations emerged at higher velocities. This discrepancy, which is evident in the velocity (Fig.~\ref{fig:velocity_profile}(a) and (b)) and shear rate distributions (Fig.~\ref{fig:velocity_profile} (c)), indicates that either non-isothermal effects play a role or that velocity gradient components beyond axial shear make significant contributions. To examine the plausibility of the experimental results, we compared them with the non-isothermal CFD solution. Furthermore, we analytically determined the radial velocity component of the experimental data, thereby obtaining all components of the strain rate tensor needed to investigate rheological characteristics.

\begin{figure*}
	\centering
		\includegraphics[scale=0.9]{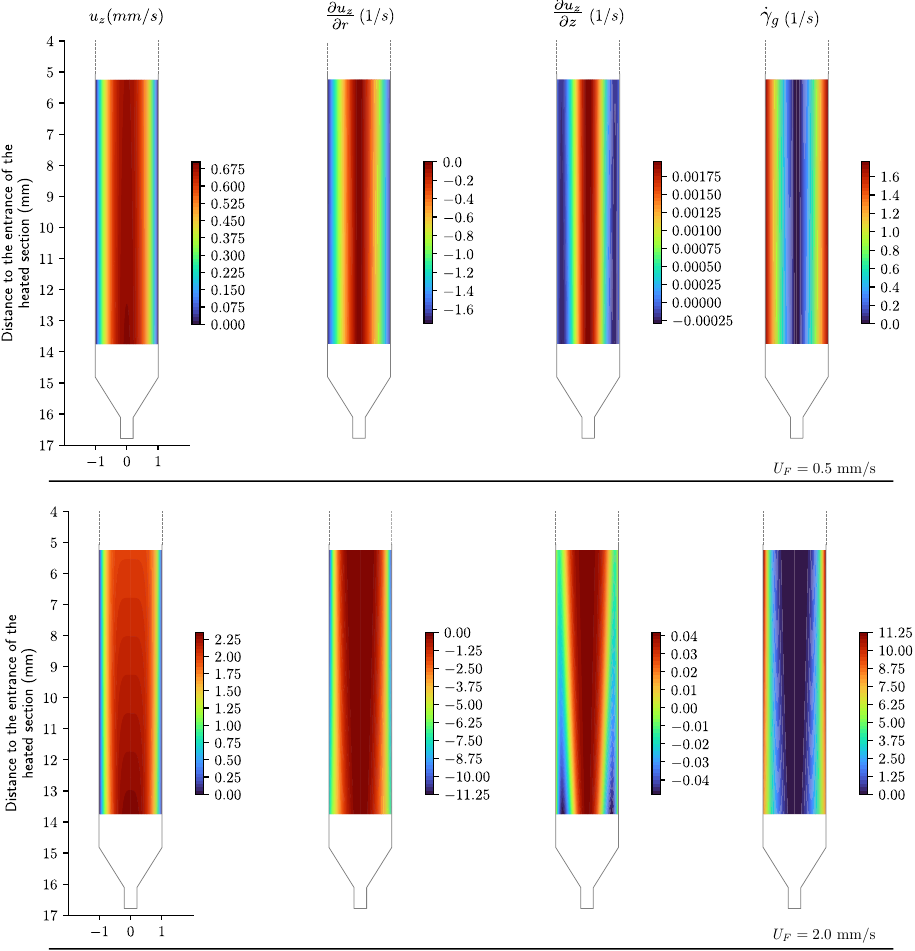}
	\caption{Color-coded visualization of the axial velocity \(u_z(r,z)\), selected velocity gradient fields, and the generalized shear rate for the lowest and highest filament velocities investigated, highlighting differences in the spatial distribution and magnitude of shear and extensional components. To enhance visibility across the wide range of filament speeds, individual color scales are applied for each case. This choice emphasizes qualitative changes in flow structure, whereas a uniform scale would suppress relevant details at lower velocities.}\protect
	\label{fig:color_coded}
\end{figure*}

We infer the radial velocity from Eq.~\ref{eq:fitting} by rearranging the continuity equation in cylindrical coordinates and integrating
\begin{equation}
    \frac{\partial u_{z}}{\partial z} + \frac{1}{r}\frac{\partial}{\partial r}(r u_{r})=0, \notag
\end{equation}
which leads to
\begin{equation}
    u_{r} = -\frac{1}{r}\left(\int r \frac{\partial u_{z}}{\partial z}\, dr  + C\right). \label{eq:ur}
\end{equation}
The integration constant \(C\) can be eliminated by applying either: i) the finiteness condition \(u_{r}<\infty\) at \(r=0\), or ii) the no-slip boundary condition \(u_{r}(r=R)=0\).  Since the continuity equation represents a first-order partial differential equation, only one boundary condition can be enforced, and one must choose the violation that is least significant and physically most acceptable. Enforcing option (ii) causes a singularity of $u_{r}\to\infty$ at the center of the flow, and likewise for its gradients$\frac{\partial u_r}{\partial r}$ and $\frac{\partial u_r}{\partial z}$, which would imply unbounded stresses on the centerline. We therefore adopt option (i) and set $C=0$. The resulting unphysical wall-normal velocity at $r=R$ remains small compared with the divergence at the center. With $u_{r}$ defined, we then compute the remaining components of the infinitesimal strain-rate tensor $\dot{\boldsymbol{\epsilon}}$ by straightforward differentiation
\begin{eqnarray}
    \dot{\boldsymbol{\epsilon}} &=& \begin{bmatrix}
                      \dot{\epsilon}_{r} & \dot{\gamma}_{rz} \\
                      \dot{\gamma}_{rz} & \dot{\epsilon}_{z}
                      \end{bmatrix}, \label{eq:strainratetensor}\\
    \dot{\epsilon}_{r} &=& \frac{\partial u_{r}}{\partial r}, \\
    \dot{\epsilon}_{z} &=& \frac{\partial u_{z}}{\partial z}, \\
    \dot{\gamma}_{rz} &=& \frac{1}{2}\left(\frac{\partial u_{z}}{\partial r} + \frac{\partial u_{r}}{\partial z}\right).
\end{eqnarray}
The exact expressions can be found in the Appendix~\ref{sec:vel-grad}. Examining the absolute magnitudes of each tensor component and velocity gradient (Fig.~\ref{fig:color_coded} and for more details Fig.~S13), the flow is predominantly governed by \(\frac{\partial u_{z}}{\partial r}\), which exceeds the extensional components and \(\frac{\partial u_{r}}{\partial z}\) roughly by at least two orders of magnitude. To further quantitatively assess the deformation type, we compute the main invariants of the strain rate tensor
\begin{eqnarray}
    J_{1} &=& \dot{\epsilon}_{I} + \dot{\epsilon}_{II}, \label{eq:j1}\\
    J_{2} &=& \dot{\epsilon}_{I}^{2} + \dot{\epsilon}_{II}^{2} \label{eq:j2},
\end{eqnarray}
using the principal strain rates \(\dot{\epsilon}_{I},\dot{\epsilon}_{II}\) which are the eigenvalues of \(\boldsymbol{\dot{\epsilon}}\). The near-zero magnitude of \(J_{1}\) and magnitude analysis (Fig.~S14) indicate predominantly shear-dominated flow conditions throughout, except near the center. To evaluate whether significant shear-thinning occurs under isothermal conditions, we calculate the generalized shear rate \cite{agassant2017polymer}
\begin{equation}
    \dot{\gamma}_{g} = 2\sqrt{J_{2}},
\end{equation}
which remains consistently below the critical shear rate within the barrel section for all measured filament velocities, apart from a thin layer adjacent to the wall at the highest filament velocity (Fig. \ref{fig:color_coded} ). In conjunction with the one-dimensional numerical solutions of the Carreau-Yasuda fluid which also show no detectable shear thinning for all different filament speeds (Fig. \ref{fig:velocity_profile}(a)), we exclude shear-thinning behavior under isothermal flow assumptions as a possible cause for the observed non-linearity. 

To assess whether elastic effects could account for the observed flow field, we examine the relaxation spectrum (Fig.~S3, Table~S1). The longest relaxation time contributes only \(0.2\%\) to the total material response, and the second-longest relaxation time is close to the inverse of the critical shear rate \(\dot{\gamma}_{c}\), defined by the intersection between the Newtonian plateau and the power-law regime (Fig.~\ref{fig:master}, Table~\ref{Rheo_parameters}). We therefore use \(\dot{\gamma}_{c}\) as a benchmark for elastic relevance. Comparison with the measured generalized shear rate in Fig.~S14 shows that relaxation dynamics are substantially faster than the imposed deformation everywhere except, at the highest filament velocity, within a thin near-wall layer. Nevertheless, deviations from isothermal behavior already appear at a filament velocity of \SI[per-mode=symbol]{1.0}{\milli\meter\per\second}, indicating that elastic effects are secondary.


Consequently, we attribute the observed deviations from ideal pipe-flow predictions to non-isothermal effects. This interpretation is consistent with the CFD results. 
At the lowest filament velocity, the polymer has sufficient time to reach thermal equilibrium (Fig.~S7), which explains the absence of any noticeable variation in the velocity profile along the flow direction. In this case, the CFD results show excellent agreement with the experimental velocity and shear-rate profiles (Fig.~\ref{fig:velocity_profile} (b) and Fig.~S6), indicating that the simulations accurately capture the underlying flow physics. With increasing filament velocity, the CFD results indicate that the polymer no longer has sufficient residence time to reach thermal equilibrium. This leads to generally flatter velocity profiles that gradually lose their flatness along the flow direction. In this regime, the CFD results still reproduce the experimental data qualitatively well. These findings not only confirm the predictive capability of the numerical model but also support our interpretation that the nozzle flow is shear-dominated and strongly influenced by non-isothermal effects.


\subsection{Temperature field approximation}
Direct measurement of the temperature inside the nozzle is not feasible without significantly disturbing flow conditions, such as when introducing a thermal sensor into the filament \cite{peng_complex_2018}. Even if a temperature sensor were employed, it would provide only a single-point measurement. To illustrate a possible approach to overcome this limitation, we present a method to infer an approximate temperature field directly from the measured velocity fields. We simplify the flow to a purely pressure-driven shear flow, an assumption supported by the observed velocity gradients (Fig.~S13) and further corroborated by our simulations, which show that the pressure distribution exhibits only a weak radial dependence (Fig.~S9). Under this assumption we obtain \(\dot{\gamma}_{g}\approx \frac{\partial u_{z}}{\partial r}\), reducing the momentum equation to the well-known Poiseuille flow in a pipe. The shear stress in pipe flow is
\begin{equation}
    \tau_{rz} \approx  \frac{\partial p}{\partial z} \frac{r}{2},
\end{equation}
where the pressure gradient \(\frac{\partial p}{\partial z}\) is obtained by assuming a known wall temperature \(T=T_{h}\) and utilizing the measured wall shear rate \(\dot{\gamma}_{w}\left(z\right)=\frac{\partial u_{z}}{\partial r}\left(R,z\right)\). For the Carreau-Yasuda fluid, the true wall viscosity is \(\eta_{w} = \eta \left(T_{h},\dot{\gamma}_{w}\right)\), allowing calculation of the pressure gradient
\begin{equation}
    \frac{\partial p}{\partial z} = \eta_{w}\dot{\gamma}_{w} \frac{2}{R}.
\end{equation}
The \textit{apparent} viscosity distribution within the barrel section can thus be defined by the standard relationship \(\eta_{a} \left(r,z\right) = \frac{\tau_{rz}}{\dot{\gamma}}\), resulting in
\begin{equation}
    \eta_{a} \left(r,z\right) = \eta_{w} \frac{r}{R} \frac{\dot{\gamma}_{w}}{\dot{\gamma}\left(r,z\right)}.
\end{equation}

To infer the actual viscosity, we incorporate the time-temperature superposition via the shift factor \(a_{T}\) into the Carreau-Yasuda viscosity model, yielding
\begin{equation}
    \left( \frac{\eta_{a}}{\eta_{0}}  \right)^{\frac{n_{1}}{n_{2}-1}} = \left[\left(\lambda\dot{\gamma}\right)^{n_{1}} a_{T}^{-n_{1}} + 1\right]a_{T}^{\frac{n_{1}}{n_{2}-1}}, \label{eq:at_transc}
\end{equation}
which, when set equal to \(\eta_{a}\), becomes a transcendental equation in terms of \(a_{T}\). Although no exact analytical solution exists, we can derive two asymptotic solutions:
i) for \(\left(\lambda\dot{\gamma}\right)^{n_{1}} \ll 1\) (Newtonian limit)
\begin{eqnarray}
    a_{T} \approx \frac{\eta_{a}}{\eta_{0}}.
\end{eqnarray}
ii) for \(\left(\lambda\dot{\gamma}\right)^{n_{1}} \gg 1\) (Power-law limit)
\begin{equation}
    a_{T} \approx \left(\lambda\dot{\gamma}\right)^{\frac{1-n_{2}}{2-n_{2}}} \left(\frac{\eta_{a}}{\eta_{0}}\right)^{\frac{1}{2-n_{2}}}.
\end{equation}
We numerically solve Eq. \ref{eq:at_transc}, using the Newtonian solution as an initial guess, reverting to the power-law limit if convergence is not achieved. The temperature is then recovered by inverting the shift factor-temperature relation in Eq. \ref{eq:at}
\begin{eqnarray}
    T = \frac{\log{\left(a_{T}\right)} \left(T_{0}-C_{2}\right) + C_{1}T_{0} }{C_{1}+\log{\left(a_{T}\right)}}.
\end{eqnarray}

\begin{figure}
\includegraphics[scale=1.1]{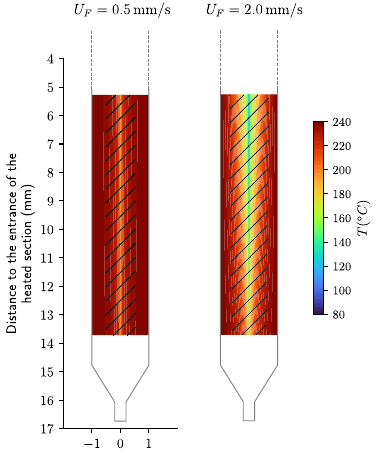}
\caption{\label{fig:temp} Approximated temperature fields derived from the obtained velocity profiles at the lowest and highest filament velocities, illustrating significant temperature gradients developing towards the centerline of the nozzle. The black line bounds the region of low confidence as determined by Eq. \ref{eq:error-estimate}.}
\end{figure}

Because the CFD solution reproduces the experimental trends for both the dependence of the velocity field on filament velocity and the effect of heater temperature, we conclude that the model captures the relevant physics and provides a reasonable basis for interpreting the measurements (Additional CFD results are provided in the Supplementary Information~S5). We therefore use the CFD data as an independent consistency check for the reconstructed temperature field, as shown in Fig.~\ref{fig:velocity_profile}(d) and Fig.~\ref{fig:temp}.

The reconstructed temperature field and the CFD solution agree well near the nozzle wall. Toward the center, deviations appear where the reconstruction develops a singularity that arises from the chosen fitting function (Fig.~\ref{fig:velocity_profile}(d)). Agreement within approximately half the nozzle radius improves with increasing filament velocity and reaches a precision of \(\SI{10}{\kelvin}\) to \(\SI{15}{\kelvin}\). Closer to the center, the singularity leads to large discrepancies. Several factors contribute to the remaining differences in addition to the singularity: (i) artifacts of the selected fitting function distort the temperature trend, and (ii) the CFD assumes ideal heat transfer between wall and melt, which likely overestimates the actual heat flux.


The singularity can be rationalized as follows. The shear stress varies linearly with radius \(r\), whereas the shear rate exponent \(b-1>1\) (Fig.~S5) leads to diverging apparent viscosity as \(r\to0\). Regularized spline fits or low-order polynomials can remove the formal divergence, yet because the current fitting function reproduces the data well, such alternatives would still yield very large, though finite, viscosities near the center. Two explanations are plausible: either the Newtonian regime at the center is smaller than the spatial resolution of the measurement \(\approx \SI{40}{\micro\meter}\), or the velocity gradients there fall below the detection limit of the present technique. The analysis in Subsection~\ref{sec:exp-lim} supports the latter as the more likely cause.


\subsection{Experimental limitations \label{sec:exp-lim}}

The data presented in this study is primarily subject to two sources of experimental uncertainty: (i) inaccuracies arising from the feeding mechanism of the extrusion setup, and (ii) uncertainties associated with the X-ray particle tracking method and its parameter choices. 

\begin{figure}
	\centering
	\includegraphics[width=0.48\textwidth]{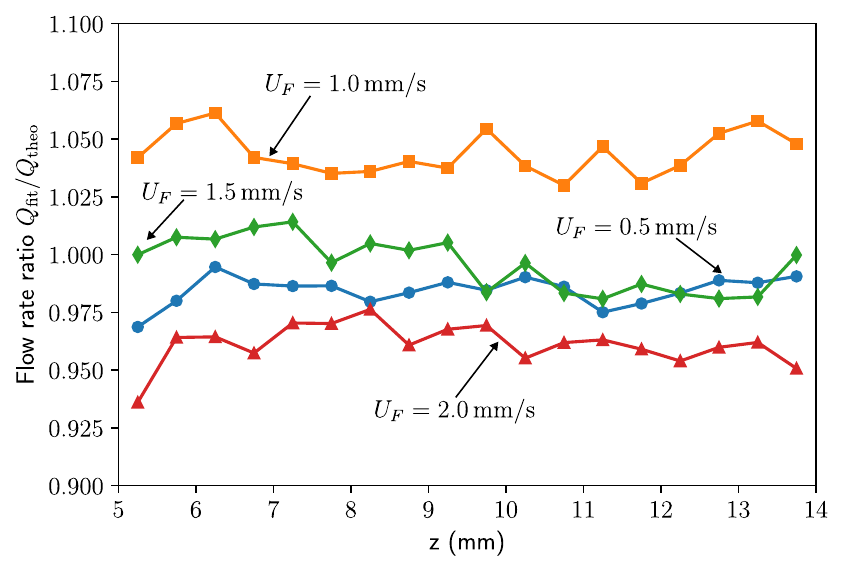}
	\caption{Comparison of volumetric flow rates from gravimetry and XPTV. Plotted is the ratio $Q_{\mathrm{grav}}/Q_{\mathrm{fit}}$ for each filament velocity, where $Q_{\mathrm{fit}}$ is obtained by integrating the fitted velocity profiles $u_z(r)$ from Eq.~\ref{eq:fitting}. A value of 1 indicates perfect agreement.}
    \protect
	\label{flowrate}
\end{figure}

To evaluate whether the feeding mechanism introduces a systematic bias, following Ref.~\onlinecite{coogan_-line_2019} we compare the volumetric flow rate inferred from our curve-fitting procedure with an independent gravimetric measurement. Specifically, the extrudate mass collected over a three-minute interval was recorded and converted to a volumetric throughput ($Q_{\mathrm{grav}}$) for comparison with the fitted flow rate ($Q_{\mathrm{fit}}$). Fig.~S10 demonstrates that the calibration was adequate, with the feeding mechanism reaching a filament velocity near the setpoint. Investigating the ratio of volumetric flow rates $Q_{\mathrm{grav}}/Q_{\mathrm{fit}}$ shown in Fig.~\ref{flowrate} reveals that the curve-fitting approach yields lower values, except for the \SI[per-mode=symbol]{1.0}{\milli\meter\per\second} condition. The largest discrepancy was observed at the \SI[per-mode=symbol]{2.0}{\milli\meter\per\second}, amounting to 5\%. While this discrepancy is not negligible, it is sufficiently small that it does not compromise the conclusions of this study. 

A plausible origin is a speed-dependent detection bias in the particle-velocity estimation: as particle speed increases, the probability of correctly determining velocities decreases, leading to an under-detection of the fastest particles at \SI[per-mode=symbol]{2.0}{\milli\meter\per\second}  and thereby biasing the envelope fit toward lower apparent flow rates. This bias is likely to occur specifically at this condition, since the corresponding velocities approach the upper limit of what could be resolved with the available X-ray imaging system.

Furthermore, it is important to note that the number of data points within the evaluation window directly governs the robustness of the envelope fit. Confidence and accuracy improve when more data points are available, for example by increasing tracer concentration or enlarging the spatial averaging window $\Delta z$ to include more data in the evaluation. To preserve spatial resolution, we therefore use a narrow $\Delta z$ and accept higher variance as the trade-off. 

To assess uncertainties associated with the X-ray particle tracking method and its parameter choices, first, we mirrored the data points previously used for fitting the power-law velocity distributions (Fig.\ref{fig:particle_paths}) using a spline fit and calculated the magnitude of the difference between mirrored and original data sets as an indirect measure of symmetry (Fig.~S11). It is important to note, however, that deviations identified through this method do not exclusively represent measurement errors, as actual asymmetric flow phenomena may also exist due to imperfect alignment of the filament with the barrel. Instead, this analysis quantifies the total deviation from ideal behavior resulting from the FFF process, finite sampling times, and visualization inaccuracies. This approach yielded an average relative difference of approximately \(5\%\), with higher deviations observed near the nozzle boundaries, where the transmission path is narrower, thus reducing particle detectability. 

Based on this relative error, we can estimate regions with reliable velocity gradients and, by extension, more reliable reconstructed temperature fields. Suppose there is an absolute error $\Delta u_{z}$ in determining the axial velocity $u_{z}$. Two successive bin values $u_{i}$ and $u_{i-1}$ can only be reliably distinguished if their difference exceeds this error,
\begin{equation}
    \Delta u_{z} \leq \left| u_{i} - u_{i-1} \right|. 
\end{equation}
Dividing by the spatial resolution $h$ which we here take to be the bin width $h = 40\,\mu\mathrm{m}$) gives
\begin{equation} \label{error_eq}
    \frac{\Delta u_{z}}{h} \leq \frac{|u_{i} - u_{i-1}|}{h}. 
\end{equation}
From Fig.~S11 we estimate an average relative error $\delta \approx 5\%$ and set $\Delta u_{z} = \delta\, u$. Interpreting the right-hand side of Eq.~\ref{error_eq} as a finite-difference approximation leads to the resolvability condition
\begin{equation} 
   \frac{ u \delta}{h} \leq \frac{\left| u_{i} - u_{i-1} \right|}{h} \approx \left| \frac{\partial u_{z}}{\partial r} \right|. \label{eq:error-estimate} 
\end{equation}
In other words, the local velocity gradient must exceed a threshold set by the relative velocity error and the spatial resolution to be reliably resolved. Regions where this condition is not satisfied for the temperature are masked in Fig.~\ref{fig:temp}. Comparing the region of largest disagreement between CFD and the reconstructed temperature in Fig.~\ref{fig:velocity_profile}(d) (approximately \(x\geq \frac{R}{2} \)) with the masked regions in Fig.~\ref{fig:temp} (approximately \(x\geq \frac{R}{2} \)) shows that these regions coincide. This suggests that the fitting function was chosen in parts due to measurement limitations in the velocity resolution, which then contribute to the temperature divergence toward the center observed in Fig.~\ref{fig:velocity_profile}(d) and Fig.~\ref{fig:temp}. The proposed approach enables reconstruction of spatial temperature distributions that are inaccessible to conventional measurement techniques. Nevertheless, the present results should be regarded as a case study and considered preliminary, since certain regions cannot yet provide reliable temperature values.

The maximum geometric magnification error of approximately \(1.3\%\) at the barrel walls is significantly smaller than the uncertainties seen in the velocity profiles, making it an unlikely dominant source of error. Moreover, the small Stokes numbers imply negligible particle inertia, consistent with the observation that the largest measurement uncertainties occur at the lowest filament velocities.

\section{Conclusion and outlook}
We present the first application of X-ray Particle Tracking Velocimetry (XPTV) to analyze polymer melt flows in optically opaque environments. By embedding tungsten powder as tracer particles into polymer filaments, we effectively characterized the internal flow behavior within a Fused Filament Fabrication (FFF) printer nozzle. This approach allowed us to make rheological assessments of an industrially relevant process. 

We resolved velocity profiles with estimated relative accuracies of 5\% to 10\%. Our results revealed deviations from Newtonian behavior that could not be detected from apparent shear rate analyses alone, strongly indicating heterogeneous, non-isothermal flow conditions within the barrel section of the nozzle. A key strength of the method is that, by enforcing continuity, we infer the radial velocity component from the measured axial field, which allows us to assemble the full strain-rate tensor even in a regime where axial motion dominates and the radial component is intrinsically very small and below practical detectability of radial particle motion. The resulting tensor shows that the flow is predominantly shear-dominated, with minor extensional components concentrated near the centerline. We also reconstructed an estimate of the temperature field in the barrel. Its accuracy is limited by spatial resolution. Even so, the qualitative features are consistent with our CFD simulations, which were designed in line with advanced approaches reported in the literature \cite{serdeczny_numerical_2020, marion_first_2023}, and they align with prior studies \cite{kattinger2023analysis}. In particular, the results indicate that non-isothermal flow should be expected at filament velocities of about \SI[per-mode=symbol]{1}{\milli\meter\per\second}, whereas earlier expectations placed the onset only at substantially higher speeds, around \SI[per-mode=symbol]{4}{\milli\meter\per\second} and above \cite{kattinger2023analysis}. These findings support the view that the hot end limits the achievable print speed \cite{go_rate_2017, phan_rheological_2018}. The presented temperature-field estimation approach can offer clear advantages over single-point sensing with a nozzle-embedded sensor, which provides only a local measurement and cannot recover a spatial temperature distribution.

A limitation of the present study is the maximum resolvable particle velocity of the laboratory microfocus setup, which prevents measurements in the conical acceleration zone and in the capillary. Resolving these regions would require substantially shorter exposure times and higher photon flux while maintaining a small source spot size for spatial resolution, a combination that is beyond the operating envelope of our source and detector. Even synchrotron illumination is only a partial remedy due to the very small field of view in typical beamline geometries \cite{rosen_synchrotron_2024}. A second limitation is the inability to capture transient flow phenomena. The Eulerian reconstruction requires time averaging of Lagrangian particle tracks within each axial section to obtain a statistically converged distribution. 

The path forward is twofold. On the hardware side, pairing high-brightness microfocus sources with high-speed cameras is promising. Recent liquid metal jet tubes now make frame rates on the order of hundreds of frames per second feasible at useful penetration for metal nozzles \cite{parker2024lab}, which could extend measurements into the conical and capillary regions and open the door to time-resolved studies. On the analysis side, advanced postprocessing can further improve robustness: regularized spline fitting, uncertainty-aware reconstructions, and data assimilation with computational fluid dynamics (CFD) will sharpen velocity and temperature estimates. Future research directions should include the investigation of other rheologically significant flows, such as standard configurations (e.g., plate-plate Couette and Poiseuille flows) and more complex geometries (e.g., contraction and cross-slot flows), potentially involving multiphase systems. In addition, the ability to capture transient behavior, such as start/stop events and rapid feed-rate changes, should be investigated as an important step toward practical FFF applications. It is also worth investigating whether the velocity component normal to the main flow can be inferred directly from particle tracking. Finally, extending to three-dimensional or four-dimensional visualization with a single source and detector using controlled sinusoidal motion of either the imaging system or the sample is a realistic avenue to reduce system complexity and cost.

Overall, this proof-of-concept establishes XPTV as a viable tool for interrogating polymer melt flows in opaque metal environments, delivering information that is difficult to obtain by other means while defining clear, achievable routes to higher velocities, transient operation, and richer reconstructions.

\section*{Supplementary Material}
Supplementary Material is available to complement this article.  
It contains details on tracer particle incorporation (Sec.~S1), 
rheological data and model fits including WLF shift factors, Cox–Merz comparison, and generalized Maxwell spectra (Sec.~S2), 
and an analysis of trajectory processing artifacts with ROI size and TrackMate parameters (Sec.~S3).  
The one-dimensional Carreau–Yasuda solution and velocity and shear-rate profiles are provided in Sec.~S4.  
Computational fluid dynamics (Sec.~S5) includes numerical details, material properties, domain representation, and results such as non-isothermal temperature fields, transient evolution, and conjugate heat transfer.  
Further sections report error estimation and flow-rate verification, contour plots of velocity gradients and invariants, and scatter plots of the raw particle-tracking data (Sec. S8).  

\begin{acknowledgments}
This research was funded by the Deutsche Forschungsgemeinschaft (DFG, German Research Foundation) under grant number 545960701. SH acknowledges the support by the Humboldt foundation through the Feodor-Lynen fellowship and thanks his host Stefano Zapperi for his advice and support.
\end{acknowledgments}

\section*{Data Availability Statement}
The data that support the findings of this study are openly available in DataVerse at https://doi.org/10.18419/DARUS-4974.
The repository includes: (i) raw X-ray radiography as 16-bit per-frame image sequences with acquisition metadata; (ii) particle-tracking outputs.

\appendix*
\section{Equations for Velocity Gradients and Strain Rate Tensor \label{sec:vel-grad}}
The fit parameters \(a\) and \(b\) can be approximated by linear functions
\begin{align}
    a\left(z\right) = \alpha_{0}z + \alpha_{1} \\
    b\left(z\right) = \beta_{0}z + \beta_{1}
\end{align}
By substituting these expressions into Eq.~\ref{eq:fitting} and subsequently solving Eq.~\ref{eq:ur}, an approximate analytical expression for the radial velocity \(u_{r,z}\) can be derived
\begin{widetext}
\begin{align}
    u_{r}(r, z) =& \left(-1\right) \left\{ \alpha_{0} r + \frac{r^{\beta_{0}z + \beta_{1}+1}}{(b(z)+2) R^{\beta_{0}z + \beta_{1}}} \right. \left. \left[ \left( \alpha_{0}z + \alpha_{1} \right) \beta_{0} \left( \ln\left(\frac{r}{R}\right) - \frac{1}{\beta_{0}z + \beta_{1}+2} \right) - \alpha_{0} \right] \right\}.
\end{align}
The velocity gradients are subsequently determined by differentiating \(u_{r}\) and \(u_{z}\) with respect to the appropriate spatial coordinates
\begin{align}
    \frac{\partial u_{z}}{\partial z}\left(r,z\right) = \alpha_{0} \left[1-\left(\frac{r}{R}\right)^{\beta_{0}z+\beta_{1}}\right] + \beta_{0} \left( \alpha_{0}z + \alpha_{1} \right) \left( \frac{r}{R}\right)^{\beta_{0}z+\beta_{1}} \ln{\left( \frac{r}{R} \right)}
\end{align}
\begin{align}
    \frac{\partial u_{r}}{\partial r}(r, z) =& \left(-1\right) \left\{ \alpha_{0} + \frac{r^{\beta_{0}z + \beta_{1}} (\beta_{0}z + \beta_{1}+1)}{(\beta_{0}z + \beta_{1}+2) R^{\beta_{0}z + \beta_{1}}} \right. \left. \left[ \beta_{0} \left( \alpha_{0} z + \alpha_{1} \right) \left( \ln\left(\frac{r}{R}\right) + \frac{1}{\beta_{0}z + \beta_{1}+1} - \frac{1}{\beta_{0}z + \beta_{1}+2} \right) - \alpha_{0} \right] \right\}
\end{align}
\begin{equation}
    \frac{\partial u_z}{\partial r}(r, z) = \left(-1\right) \left( \alpha_{0}z + \alpha_{1} \right)\left( \beta_{0}z + \beta_{1} \right) \frac{r^{\beta_{0}z + \beta_{1}-1}}{R^{\beta_{0}z + \beta_{1}}}
\end{equation}

\begin{align}
    \frac{\partial u_r}{\partial z}(r, z) =& \left(-1\right) \frac{r^{\beta_{0}z + \beta_{1}+1}}{(\beta_{0}z + \beta_{1}+2) R^{\beta_{0}z + \beta_{1}}}  \beta_{0} \left( \ln\left(\frac{r}{R}\right) - \frac{1}{\beta_{0}z + \beta_{1}+2} \right) \notag \\ & \left\{ \alpha_{0} +  \beta_{0} \left[ \left(\alpha_{0}z + \alpha_{1}\right) \beta_{0} \left( \ln\left(\frac{r}{R}\right) - \frac{1}{\beta_{0}z + \beta_{1}+2} \right) - \alpha_{0} \right] \right\} 
\end{align}
\end{widetext}
These velocity gradients subsequently define the components of the strain rate tensor as follows
\begin{equation}
    \dot{\epsilon}_{r} = \frac{\partial u_{r}}{\partial r}
\end{equation}
\begin{equation}
    \dot{\epsilon}_{z} = \frac{\partial u_{z}}{\partial z}
\end{equation}
\begin{equation}
    \dot{\gamma}_{rz} = \frac{1}{2} \left( \frac{\partial u_{z}}{\partial r} + \frac{\partial u_{r}}{\partial z} \right)
\end{equation}

\section*{References}
\bibliography{aipsamp}

\end{document}